\documentclass[10pt,twocolumn,english,prd,superscriptaddress,nofootinbib,preprintnumbers,showpacs,floatfix]{revtex4-2}

\usepackage[utf8]{inputenc}
\usepackage{amsmath}
\usepackage{amsfonts}
\usepackage{amssymb}
\usepackage{graphicx} 
\usepackage[caption=false]{subfig}
\graphicspath{ {figures/} }
\usepackage[usenames,dvipsnames]{xcolor}
\usepackage{hyperref}   
\hypersetup{
    colorlinks=true, 
    citecolor=MidnightBlue,
    linkcolor=MidnightBlue,
    urlcolor=Cyan}
\usepackage{tikz}
\usepackage{verbatim} 
\usepackage{soul} 
\definecolor{purple}{rgb}{1,0,1}
\definecolor{lime}{HTML}{A6CE39} 

\begin{document}

\title{On the possibility of emergent light cones from relational shape dynamics}

\author{Francisco S. N. Lobo} 
\email{fslobo@ciencias.ulisboa.pt}
\affiliation{Instituto de Astrof\'{i}sica e Ci\^{e}ncias do Espa\c{c}o, Faculdade de Ci\^{e}ncias da Universidade de Lisboa, Edifício C8, Campo Grande, P-1749-016 Lisbon, Portugal}
\affiliation{Departamento de F\'{i}sica, Faculdade de Ci\^{e}ncias da Universidade de Lisboa, Edif\'{i}cio C8, Campo Grande, P-1749-016 Lisbon, Portugal}


\date{\LaTeX-ed \today}
\begin{abstract}

We show that a universal propagation scale can emerge from purely relational, scale-invariant $N$-body dynamics formulated on shape space, i.e.~the space of configurations modulo translations, rotations, and dilatations. Although pure shape dynamics treats only unparametrized curves as fundamental, we adopt an affine parametrization as a gauge choice to perform a perturbative analysis, with all physically meaningful results expressed in parametrization-independent terms. 
Linear perturbations around central configurations satisfy second-order equations on shape space whose high-frequency spectrum defines a dimensionless constant $c_{\mathrm{rel}}$. Under general conditions of reparametrization invariance, spectral universality, and strict hyperbolicity, $c_{\mathrm{rel}}$ functions as an emergent light-cone velocity, endowing the product space $\mathbb{R} \times \mathcal{S}$ with an effective Lorentzian structure. These results suggest that causal structure and a maximal signal speed may arise dynamically from the geometry and spectral properties of relational configuration space, providing a novel perspective on the origin of relativistic kinematics.

\end{abstract}
\maketitle
\def\HMS{{\scriptscriptstyle{\rm HMS}}}


\section{Introduction}

Relational approaches to mechanics eliminate absolute structures such as external space, orientation, and overall scale \cite{Mercati:2018ufj,Barbour:2014bga}, retaining only those features of a physical system that are invariant under global transformations. In such frameworks, the primitive ontology consists not of particles embedded in a fixed background geometry, but of mutual relations among particles \cite{Mercati:2014ama,Gomes:2012hq}. Consequently, only dimensionless ratios, such as relative separations normalized by a relational length scale, or angular configurations among triples of particles, carry physical meaning. Absolute position, absolute orientation, and absolute size are treated as gauge redundancies, in close analogy with the treatment of gauge degrees of freedom in constrained Hamiltonian systems \cite{Lanczos:1970}.

A natural and mathematically precise arena for implementing this program is provided by the classical $N$-body problem \cite{Mercati:2018ufj,Arnold:1989}. Its full configuration space, $Q \cong \mathbb{R}^{3N}$, contains redundant degrees of freedom associated with the action of the similarity group, comprising global translations, rotations, and uniform dilatations. By quotienting out these symmetries one obtains the \emph{shape space} $\mathcal{S}$, whose points correspond to pure shapes \cite{Barbour:2011dn,Barbour:1982gha}, i.e.\ equivalence classes of configurations differing only by similarity transformations \cite{Koslowski:2021aga,Sloan:2018lim}. The geometric and dynamical properties of such reduced configuration spaces have been extensively studied in relational particle models and geometrodynamics \cite{Barbour:1994ri}.

Indeed, dynamics on $\mathcal{S}$ is intrinsically relational and scale-invariant.~With scale eliminated, no absolute length remains, and all physically meaningful quantities must be expressed as ratios constructed from the configuration itself. Time is not externally imposed but emerges from change \cite{Barbour:1994ri}: the dynamics may be derived from a reparametrization-invariant Jacobi-type action, so that the evolution parameter carries no direct physical significance.~Physical predictions depend only on the unparametrized curve traced in $\mathcal{S}$, not on any particular choice of parameter along that curve.~This feature reflects the Machian and background-independent character of the theory \cite{Barbour:2000ad,Anderson:2012vk}.

In its strict formulation, known as Pure Shape Dynamics (PSD), the fundamental object is an unparametrized curve in shape space \cite{Koslowski:2021aga}. However, for perturbative and spectral analysis it is technically convenient to introduce an affine parameter $\lambda$ along relational trajectories. Here, we adopt such a parametrization when deriving the linearized equations and analyzing their principal part.~Indeed, all physically meaningful conclusions are expressed in parametrization-independent terms, where the relevant structures are obtained from the principal symbol of the linearized operator, which is invariant under smooth monotonic reparametrizations of $\lambda$.~Thus, the perturbative framework used here remains consistent with the unparametrized ontology of PSD, even though intermediate calculations employ a convenient clock variable.

The central question addressed in this work is whether a quantity analogous to the speed of light can arise purely from relational, scale-invariant dynamics, without presupposing Lorentz invariance, a background spacetime, or an a priori causal structure. In conventional physics, the speed of light $c$ appears as a universal constant determining the invariant causal structure of spacetime \cite{Einstein:1905ve}. It enters the spacetime metric $ds^2 = -c^2 dt^2 + d\mathbf{x}^2$ and fixes the maximal propagation speed of signals, as encoded in the light-cone structure of special relativity \cite{Wald:1984rg}.

In a relational, scale-free framework, no background metric or fundamental conversion factor between spatial and temporal units is assumed. If an analogue of $c$ is to emerge, it must arise dynamically from the intrinsic geometry of shape space and from the structure of the relational equations of motion. More precisely, one must identify: (i) a relational notion of spatial distance encoded in the metric on $\mathcal{S}$; (ii) an emergent temporal parameter reconstructed from change; and (iii) a universal characteristic speed governing the high-frequency propagation of perturbations. This propagation speed is determined not by coordinate choices, but by the hyperbolic principal symbol of the linearized operator, whose null cone, if it exists, defines the maximal propagation directions in the geometric optics limit \cite{Courant:1962}.

We show that, at the level of linear perturbation theory about distinguished configurations (such as central configurations) \cite{Battye:2002gn,saari_cc}, the relational equations admit a natural definition of a dimensionless characteristic speed. This quantity is determined by the intrinsic metric on shape space together with the Hessian of the scale-invariant potential evaluated at the background configuration. In an appropriate continuum or large-$N$ regime, the principal part of the perturbation operator becomes hyperbolic and defines, up to conformal rescaling associated with clock redefinitions, a Lorentzian quadratic form on $\mathbb{R}_\lambda \times \mathcal{S}$. The associated characteristic surfaces then determine the maximal propagation speed of high-frequency relational disturbances.

In this way, an analogue of the speed of light appears not as a fundamental constant postulated at the outset, but as an emergent property of the relational dynamics. Its value is fixed entirely by internal, dimensionless structures of the theory.~No claim is made that Lorentzian signature is fundamental; rather, it arises, within a well-defined perturbative regime, from the hyperbolic structure of the linearized relational equations. The remainder of this work develops this construction in detail and analyzes the precise conditions under which such an emergent characteristic speed can consistently be interpreted as defining an effective causal structure.

This work is organized as follows. Section~\ref{SecII} introduces relational configuration space and defines shape space. Section~\ref{SecIII} presents the reparametrization-invariant relational action and the emergent temporal parameter. In Sec.~\ref{SecIV}, we construct the scale-invariant potential and identify central configurations as distinguished backgrounds. Section~\ref{SecV} develops linear perturbation theory on shape space, and Sec.~\ref{SecVI} defines a conditional emergent propagation parameter using continuum and eikonal approximations. Section~\ref{SecVII} analyzes the emergence of an effective light-cone structure. Finally, Sec.~\ref{SecConclusion} summarizes the results and outlines possible future lines of research.

\section{Relational Configuration Space and Shape Space}\label{SecII}

Let $\mathbf{q}_i \in \mathbb{R}^3$, $i=1,\dots,N$, denote the positions of $N$ point particles with masses $m_i$. The ordinary configuration space of the system is $Q \;\cong\; \mathbb{R}^{3N}$, whose points represent ordered $N$-tuples of particle positions. This space contains substantial redundancy from the relational point of view: two configurations that differ only by a global translation, rotation, or overall rescaling represent the same physical situation if only relative structure is regarded as meaningful \cite{Barbour:1982gha,Barbour:2011dn}.

The similarity group acting on $Q$ is $\mathrm{Sim}(3) = \mathbb{R}^3 \rtimes
\bigl( {SO}(3) \times \mathbb{R}_+ \bigr)$, where $\mathbb{R}^3$ generates translations, $SO(3)$ rotations, and $\mathbb{R}^+$ uniform dilatations. Its action on configurations is given by: $\mathbf{q}_i \;\mapsto\; \lambda\, R\, \mathbf{q}_i + \mathbf{a}$, with $\mathbf{a} \in \mathbb{R}^3$, $R \in SO(3)$, and $\lambda > 0$.

To implement relationalism, we quotient out this action and define
\begin{equation}
	\mathcal{S} \;=\; \frac{Q}{\mathbb{R}^3 \rtimes
		\bigl( {SO}(3) \times \mathbb{R}_+ \bigr)}.
\end{equation}
The resulting space $\mathcal{S}$ is \emph{shape space}. Each point of $\mathcal{S}$ corresponds to an equivalence class of configurations differing only by global similarity transformations \cite{Barbour:2011dn}. 

For $N \ge 3$, the dimension of $\mathcal{S}$ is $\dim \mathcal{S} = 3N - 7$, reflecting the removal of three translational, three rotational, and one dilatational degrees of freedom. The residual degrees of freedom describe pure shapes, encoded in relative separations and angles.

A convenient step is to remove translations by passing to the center-of-mass frame: 
\begin{equation}
	\mathbf{q}_{\mathrm{cm}} = \frac{1}{M} \sum_{i=1}^{N} m_i \mathbf{q}_i,
	\qquad
	M = \sum_{i=1}^{N} m_i.
\end{equation}
Define relative coordinates: $\mathbf{r}_i = \mathbf{q}_i - \mathbf{q}_{\mathrm{cm}}$ and $\sum_i m_i \mathbf{r}_i = 0$.
The translational redundancy is thereby removed, and the configuration now lives in a $(3N-3)$-dimensional linear subspace \cite{Arnold:1989}.

The center-of-mass moment of inertia,
\begin{equation}
	I_{\mathrm{cm}}
	=
	\sum_{i=1}^{N}
	m_i
	\|\mathbf{r}_i\|^2,
	\label{CM_mom_inertia}
\end{equation}
provides a natural measure of the overall size of the configuration. Under a global scaling $\mathbf{r}_i \mapsto \lambda \mathbf{r}_i$, one has $I_{\mathrm{cm}} \mapsto \lambda^2 I_{\mathrm{cm}}$.
Thus, $\sqrt{I_{\mathrm{cm}}}$ transforms linearly under dilatations and can serve as an intrinsic length scale \cite{Barbour:1982gha}.

We define the relational length $L = \sqrt{I_{\mathrm{cm}}}$, and introduce dimensionless, scale-invariant variables $\tilde{\mathbf{r}}_i =\mathbf{r}_i/L$.
By construction, $\sum_i m_i \|\tilde{\mathbf{r}}_i\|^2 = 1$,
so the $\tilde{\mathbf{r}}_i$ lie on a unit sphere in the $(3N-3)$-dimensional relative configuration space.

After removing scale, the relative configuration space becomes a sphere $S^{3N-4}$ (for fixed masses), with antipodal points identified if reflections are considered gauge \cite{Barbour:2011dn}. Quotienting further by rotations corresponds to dividing by the $SO(3)$ action on this sphere. Thus, shape space may be viewed geometrically as
\begin{equation}
	\mathcal{S}
	\;\cong\;
	\frac{S^{3N-4}}{SO(3)}.
\end{equation}

The kinetic energy, $T = \frac{1}{2} \sum_i m_i \dot{\mathbf{r}}_i^2$, induces a natural metric on relative configuration space \cite{Arnold:1989}. After projecting orthogonally to the gauge directions (translations, rotations, and dilatations), one obtains an induced Riemannian metric $\mathcal{G}_{AB}$ on $\mathcal{S}$ \cite{Barbour:1982gha,Barbour:2011dn}. This metric endows shape space with intrinsic geometry and allows the definition of relational distances between nearby shapes:
\begin{equation}
	ds_{\mathrm{shape}}^2
	=
	\mathcal{G}_{AB}\, dq^A dq^B.
\end{equation}

Thus, shape space is not merely a quotient set but a curved Riemannian manifold with a well-defined metric structure inherited from the kinetic term. Importantly, this geometry is dimensionless: all lengths in $\mathcal{S}$ are ratios of physical lengths.

Because the dilatational degree of freedom has been eliminated, no absolute size remains, and any two configurations related by a global rescaling correspond to the same point in $\mathcal{S}$. Consequently, there is no preferred unit of length, all observables must be dimensionless, and physical evolution is represented by curves in $\mathcal{S}$.
In this way, shape space provides a mathematically precise realization of relational, scale-invariant mechanics \cite{Barbour:2011dn}.

\section{Relational Action and Time Parameter}\label{SecIII}

Having identified shape space $\mathcal{S}$ as the space of purely relational configurations, we now formulate the dynamics in a manner that does not presuppose an external time parameter. The guiding principle is reparametrization invariance: the dynamical laws should determine unparametrized trajectories in configuration space, with temporal ordering emerging from change itself \cite{Barbour:1994ri}.

Let $\mathcal{Q} = Q \setminus \Delta$ denote the collision-free sector of configuration space, where $\Delta$ is the set of total and partial collisions. On $\mathcal{Q}$ the Newtonian potential
\begin{equation}
	V_N(q)
	=
	- \sum_{i<j}
	\frac{G m_i m_j}{\|\mathbf{q}_i - \mathbf{q}_j\|},
	\label{New_potential}
\end{equation}
is smooth and homogeneous of degree $-1$ under global dilatations $q \mapsto \lambda q$.

The kinetic energy defines the mass metric on $Q$:
\begin{equation}
	T(q,\dot q)
	=
	\frac{1}{2}
	\sum_{i=1}^{N}
	m_i \dot{\mathbf{q}}_i^2,
\end{equation}
where $\dot{\mathbf{q}}_i = d\mathbf{q}_i/d\lambda$ and $\lambda$ is an arbitrary monotonic parameter along a curve in $\mathcal{Q}$.

Jacobi's principle provides a timeless formulation of Newtonian mechanics at fixed total energy $E$ \cite{Mercati:2018ufj,Lanczos:1970,Arnold:1989}. The action functional is
\begin{equation}
	S_J[q]
	=
	\int
	\sqrt{2\big(E - V_N(q)\big)\, T(q,\dot q)}
	\; d\lambda,
	\label{Jacobi_action}
\end{equation}
defined on the region where $E - V_N(q) > 0$.
The integrand is homogeneous of degree one in the velocities $\dot{\mathbf{q}}_i$. Consequently, $S_J$ is invariant under arbitrary monotonic reparametrizations $\lambda \mapsto \tilde{\lambda}(\lambda)$. The parameter $\lambda$ therefore carries no physical meaning: the theory determines only the geometric image of the trajectory in $\mathcal{Q}$.

The Euler-Lagrange equations derived from the action~\eqref{Jacobi_action} are equivalent to Newton's equations when expressed in terms of a suitably defined parameter. Introducing a time parameter $t$ along a solution by
\begin{equation}
	\frac{dt}{d\lambda}
	=
	\sqrt{\frac{T}{E - V_N}},
	\label{Jacobi_time}
\end{equation}
the equations of motion reduce to
\begin{equation}
	m_i \frac{d^2 \mathbf{q}_i}{dt^2}
	=
	- \nabla_i V_N.
\end{equation}
Thus, Newtonian time $t$ is reconstructed from relational quantities along the dynamical trajectory. It is not fundamental but arises from the interplay between kinetic and potential contributions.

Jacobi's principle admits a geometric interpretation. The action \eqref{Jacobi_action} is proportional to the arc-length functional of the conformally rescaled metric
\begin{equation}
	ds_J^2
	=
	2\big(E - V_N(q)\big)
	\,
	\mathcal{G}_{ij}\, dq^i dq^j,
\end{equation}
where $\mathcal{G}_{ij}$ is the mass metric on configuration space, $\mathcal{G}_{ij}\, dq^i dq^j = \sum_{i=1}^{N} m_i\, d\mathbf{q}_i^2$.
The physical trajectories are therefore geodesics of the Jacobi metric on $\mathcal{Q}$ restricted to the energy surface.

To obtain a purely relational formulation, one quotients $\mathcal{Q}$ by the similarity group $\mathrm{Sim}(3) = \mathbb{R}^3 \rtimes
\bigl( \mathrm{SO}(3) \times \mathbb{R}_+ \bigr)$ removing translations, rotations, and global scale. The resulting space is shape space $\mathcal{S} = \mathcal{Q}/\mathrm{Sim}(3)$.

Let $q^A$ denote local coordinates on $\mathcal{S}$, and let $\mathcal{G}_{AB}$ be the Riemannian metric induced from the mass metric after symmetry reduction. In scale-invariant variables, the dynamics reduces to geodesic motion with respect to the appropriately reduced Jacobi metric. A natural intrinsic arc-length parameter $\tau$ on shape space is defined by
\begin{equation}
	d\tau^2
	=
	2\big(E - V_{\mathrm{shape}}(q)\big)
	\,
	\mathcal{G}_{AB}\, dq^A dq^B,
\end{equation}
where $V_{\mathrm{shape}}$ is the scale-invariant potential defined previously.

Because the Jacobi Lagrangian is homogeneous of degree one in the velocities, its Legendre transform is singular. The canonical Hamiltonian therefore vanishes identically, and the dynamics is governed by the primary Hamiltonian constraint $\mathcal{H} = T + V_N - E = 0$.
This constraint generates reparametrizations of the trajectory parameter. In Dirac's terminology, it reflects the gauge freedom associated with arbitrary choices of parametrization \cite{Dirac:1964}.

There is consequently no external time variable canonically conjugate to energy. Physical observables must weakly commute with the Hamiltonian constraint, ensuring invariance under reparametrizations. The genuine degrees of freedom are relational and evolve only with respect to one another.

Since the action is reparametrization-invariant, any monotonic function along a solution curve may serve as an internal clock, at least locally. Examples include the center-of-mass moment of inertia $I_{\mathrm{cm}}$, suitably defined complexity measures, or monotonic functions of the scale-invariant potential in appropriate dynamical regimes \cite{Barbour:2014bga,Mercati:2018ufj}. 

Time, in this framework, is therefore not an externally imposed parameter but an emergent ordering derived from relational change. The fundamental object is the unparametrized curve in shape space, $q^A = q^A(\lambda)$
and temporal succession corresponds to progression along this trajectory.

This relational perspective is essential for discussing propagation phenomena. In the absence of any external temporal metric, the notion of speed cannot be defined with respect to an absolute background time, but must instead be formulated entirely in relational terms. In particular, it must be defined using relational distances measured by the metric $\mathcal{G}_{AB}$ on shape space, together with relational time parameters constructed intrinsically from change within the system itself.

Only after such structures are identified can one meaningfully address the possibility of an emergent maximal propagation rate. If perturbations propagate along shape-space trajectories at a bounded rate relative to an intrinsic arc-length parameter, that bound provides a natural candidate for an emergent analogue of the speed of light.

The next step is therefore to analyze the structure of the scale-invariant potential on $\mathcal{S}$ and to study small perturbations around its critical points within this fully relational, reparametrization-invariant framework.

\section{Scale-Invariant Potential and Central Configurations}\label{SecIV}

The elimination of overall scale from the configuration space suggests that the dynamical content of the theory should likewise be expressible in scale-invariant form. The Newtonian potential, given by Eq.~(\ref{New_potential}),
is homogeneous of degree $-1$ under global dilatations,
$\mathbf{q}_i \mapsto \lambda \mathbf{q}_i$, which implies $V_N \mapsto \lambda^{-1} V_N$.
On the other hand, the center-of-mass moment of inertia, Eq.~(\ref{CM_mom_inertia}),
is homogeneous of degree $+2$ under the same transformation. Therefore, the combination \cite{Lourenco:2026uto}
\begin{equation}
	V_{\mathrm{shape}}
	= \sqrt{I_{\mathrm{cm}}}\, V_N \,,
\end{equation}
is homogeneous of degree zero and hence invariant under global rescalings. $V_{\mathrm{shape}}$ is denoted as the {\it variety} \cite{Lourenco:2026uto,Lourenco:2026lbr}, and also referred to as {\it complexity} in \cite{Barbour:2014bga,Barbour:2024zwv}. It depends only on shape, not on absolute size. Consequently, $V_{\mathrm{shape}}$ corresponds to a well-defined function on shape space $\mathcal{S}$.

Critical points of $V_{\mathrm{shape}}$ on $\mathcal{S}$ correspond to configurations for which the shape is stationary under scale-invariant variations. Equivalently, they are configurations satisfying
\begin{equation}
	\nabla_i V_N
	=
	- \lambda\, m_i (\mathbf{q}_i - \mathbf{q}_{\mathrm{cm}}),
\end{equation}
for some constant $\lambda$, identical for all $i$. These are the classical \emph{central configurations} of the $N$-body problem \cite{Smale:1970,Saari:1980}.

To understand this condition, note that
\begin{equation}
	\nabla_i I_{\mathrm{cm}}
	=
	2 m_i (\mathbf{q}_i - \mathbf{q}_{\mathrm{cm}}).
\end{equation}
At a critical point of $V_{\mathrm{shape}}$, variations tangent to shape space must leave $V_{\mathrm{shape}}$ stationary. After eliminating translational and rotational directions, this requirement reduces to the proportionality condition above: the gravitational force on each particle must be aligned with its position vector relative to the center of mass \cite{Saari:1980}.

Physically, this implies that the acceleration of each particle is proportional to its position: $m_i \ddot{\mathbf{q}}_i = - \nabla_i V_N = \lambda\, m_i \mathbf{q}_i$.
Hence, $\ddot{\mathbf{q}}_i = \lambda\, \mathbf{q}_i$, which describes homothetic motion. The configuration expands or contracts uniformly while preserving its shape \cite{Wintner:1941}.

Since $V_{\mathrm{shape}}$ is intrinsically defined on $\mathcal S$, its critical points are geometrically distinguished features of shape space. In intrinsic coordinates $q^A$ on $\mathcal S$, central configurations satisfy
\begin{equation}
	\nabla_A V_{\mathrm{shape}} = 0,
\end{equation}
where $\nabla_A$ denotes the covariant derivative associated with the shape-space metric $\mathcal{G}_{AB}$ induced by the kinetic term \cite{Barbour:2011dn}.

The covariant Hessian
\begin{equation}
	\mathcal M_{AB}
	=
	\nabla_A \nabla_B V_{\mathrm{shape}}
\end{equation}
evaluated at a central configuration encodes the local curvature of the potential landscape in shape space and determines the linear stability properties of the configuration \cite{Moeckel:1990}. If $\mathcal{M}_{AB}$ is positive definite, the configuration is a local minimum of $V_{\mathrm{shape}}$ and hence stable under small shape perturbations. If it possesses negative eigenvalues, unstable directions exist.

Central configurations thus serve as distinguished points in shape space around which relational dynamics can be systematically analyzed \cite{Saari:1980}. Because the scale degree of freedom decouples at these configurations, they provide a natural background for studying perturbations purely in the shape directions.

Moreover, since $V_{\mathrm{shape}}$ is dimensionless, the eigenvalues of its Hessian are likewise dimensionless quantities characterizing intrinsic curvature scales of shape space. These eigenvalues will play a crucial role in determining the characteristic frequencies of small perturbations and, ultimately, in defining a candidate emergent propagation speed.

Thus, central configurations are not merely special solutions of the Newtonian $N$-body problem; in the relational, scale-invariant formulation they are critical geometric structures of shape space. Their local stability properties and associated spectra provide the natural starting point for analyzing how relational disturbances propagate and whether a universal maximal propagation rate can arise intrinsically from the geometry and dynamics of $\mathcal{S}$.

\section{Linear Perturbations in Shape Space}\label{SecV}

In Pure Shape Dynamics (PSD), the fundamental object is an 
\emph{unparametrized} curve in shape space~\cite{Koslowski:2021aga}. Any parametrization of such 
a curve is regarded as gauge, and carries no intrinsic physical meaning.~In the present analysis, however, we adopt a convenient affine 
parametrization of relational trajectories in order to perform a 
systematic perturbative expansion.~This choice is understood strictly 
as a gauge fixing of the underlying reparametrization invariance. 
All physically meaningful conclusions will ultimately be expressed in 
parametrization-independent terms.

Let $q^A$ denote local coordinates on shape space $\mathcal{S}$,~with 
$A = 1,\dots,3N-7$ for $N \ge 3$.~These coordinates parametrize the true 
relational degrees of freedom after elimination of translations, rotations, 
and overall scale \cite{Barbour:2011dn}.

Relational dynamics derived from Jacobi's principle induces equations of 
motion on $\mathcal{S}$ of the form
\begin{equation}
	\frac{D^2 q^A}{D\lambda^2}
	=
	- \mathcal{G}^{AB} \nabla_B V_{\mathrm{shape}},
	\label{eq:EOM_shape}
\end{equation}
where $\mathcal{G}_{AB}(q)$ is the induced Riemannian metric on shape space 
obtained from the kinetic term, $\nabla_B$ is the covariant derivative 
compatible with $\mathcal{G}_{AB}$, and $\lambda$ is an affine parameter 
chosen after fixing the reparametrization gauge 
\cite{Arnold:1989,Barbour:1982gha}. 
Although $\lambda$ itself has no invariant significance, the geometric 
content of Eq.~\eqref{eq:EOM_shape}, namely, the relational trajectory as 
an unparametrized curve in $(\mathcal{S},\mathcal{G})$, is independent of 
this choice.

Let $q_0^A(\lambda)$ be a solution of the full nonlinear equations 
\eqref{eq:EOM_shape}. Consider a small perturbation
\begin{equation}
	q^A(\lambda)
	=
	q_0^A(\lambda)
	+
	\delta q^A(\lambda),
	\qquad
	|\delta q^A| \ll 1.
\end{equation}
Expanding the equations of motion to first order in $\delta q^A$ yields 
the covariant perturbation equation
\begin{eqnarray}
	\frac{D^2}{D\lambda^2} \delta q^A
	&+&
	\mathcal{R}^A{}_{CBD}\,
	\dot{q}_0^C \dot{q}_0^D \,
	\delta q^B
	\nonumber\\
	&+&
	\mathcal{G}^{AC}
	\nabla_C \nabla_B V_{\mathrm{shape}}
	\Big|_{q_0}
	\,
	\delta q^B
	=
	0,
	\label{eq:perturb_covariant}
\end{eqnarray}
where $\mathcal{R}^A{}_{CBD}$ is the Riemann curvature tensor of 
$(\mathcal{S},\mathcal{G})$. This equation generalizes geodesic deviation 
to motion in a potential on a curved configuration manifold 
\cite{Synge:1960}.

A particularly simple situation arises when the background solution 
corresponds to a central configuration with constant shape, so that 
$q_0^A = \mathrm{const}$ and therefore $\dot{q}_0^A = 0$ 
\cite{Smale:1970}. 
It is important to clarify that, although a central configuration corresponds to a single point $q_\ast \in \mathcal S$, it defines an equilibrium solution of the reduced dynamics $\nabla_{\dot q}\dot q = - \mathcal{G}^{AB}\nabla_B V_{\mathrm{shape}}$, where $\nabla_{\dot q}\dot q $ denotes the covariant derivative of the vector $\dot q $ along the curve itself.
Indeed, a central configuration is a critical point of the shape potential, $\nabla_A V_{\mathrm{shape}}(q_\ast)=0$.
For the constant curve $q_0(\lambda)=q_\ast$, one has $q_0^A=\mathrm{const}$ and therefore $\dot q_0^A=0$, i.e.\ $\dot q_0$ is the zero vector in $T_{q_\ast}\mathcal S$. Hence $\nabla_{\dot q_0}\dot q_0=0$, and the equation of motion is satisfied identically. The curve $q_0(\lambda)=q_\ast$ is thus an equilibrium solution on $(\mathcal S,\mathcal{G}_{AB})$, about which linear perturbations are defined in the usual manner.

In this case, the curvature term in 
Eq.~\eqref{eq:perturb_covariant} vanishes and the perturbation equation 
reduces to
\begin{equation}
	\frac{d^2}{d\lambda^2} \delta q^A
	+
	\mathcal{M}^A{}_B \, \delta q^B
	=
	0,
	\label{eq:pert_eq_corrected}
\end{equation}
where
\begin{equation}
	\mathcal{M}^A{}_B
	=
	\mathcal{G}^{AC}
	\nabla_C \nabla_B V_{\mathrm{shape}}
	\Big|_{q_0}.
\end{equation}
Since the covariant Hessian $\nabla_A \nabla_B V_{\mathrm{shape}}$ is 
symmetric, the operator $\mathcal{M}^A{}_B$ is self-adjoint with respect 
to $\mathcal{G}_{AB}$ and therefore admits a complete set of orthonormal 
eigenvectors (at least locally) \cite{Arnold:1989}.

Let $e^A_{(k)}$ denote eigenvectors satisfying
\begin{equation}
	\mathcal{M}^A{}_B \, e^B_{(k)}
	=
	\omega_k^2 \, e^A_{(k)}.
\end{equation}
Expanding the perturbation in this basis,
\begin{equation}
	\delta q^A(\lambda)
	=
	\sum_k a_k(\lambda)\, e^A_{(k)},
\end{equation}
each mode amplitude satisfies
\begin{equation}
	\frac{d^2 a_k}{d\lambda^2}
	+
	\omega_k^2 \, a_k
	=
	0.
\end{equation}
Thus, small relational disturbances near a central configuration behave 
as harmonic oscillators with respect to the chosen affine parameter.

It is crucial to emphasize that the quantities $\omega_k$ are not 
themselves invariant observables: they depend on the parametrization. 
They characterize instead the intrinsic curvature of the shape-space 
potential relative to the kinetic metric.

Let $t$ denote an emergent relational time variable obtained by a monotonic reparametrization $t = t(\lambda)$, as permitted by Jacobi's principle \cite{Lanczos:1970}.~Under this transformation, derivatives 
rescale by the appropriate Jacobian factor, and the second-order equation 
generically acquires an additional first-derivative term reflecting the 
non-affine character of the new parameter.

For backgrounds of constant shape and suitably chosen affine gauges 
such that the Jacobian varies slowly (or is constant), this extra term 
may be neglected at leading order. The perturbation equation then takes 
the approximate form
\begin{equation}
	\frac{d^2 a_k}{dt^2}
	+
	\Omega_k^2(t)\, a_k
	\approx
	0,
\end{equation}
with
\begin{equation}
	\Omega_k^2(t)
	=
	\left( \frac{d\lambda}{dt} \right)^2
	\omega_k^2.
\end{equation}

While $\omega_k$ encode intrinsic geometric information about 
$(\mathcal{S},\mathcal{G},V_{\mathrm{shape}})$, the physically realized 
frequencies $\Omega_k$ depend on the relational clock choice. Any 
observable statement must therefore be formulated in terms of 
reparametrization-invariant structures, such as ratios of frequencies 
or asymptotic scaling relations.

In systems with large $N$, the spectrum of $\mathcal{M}^A{}_B$ may 
become dense. If, in an appropriate limit admitting an effective 
continuum description, the eigenvalues exhibit asymptotic behavior
\begin{equation}
	\omega_k^2
	\sim
	c_{\mathrm{rel}}^{\,2} k^2
	\qquad
	\text{for large } k,
\end{equation}
then the coefficient $c_{\mathrm{rel}}$ defines an intrinsic 
dimensionless propagation parameter determined entirely by the 
geometry of shape space and the curvature structure of the 
scale-invariant potential.

The linear perturbation analysis therefore identifies the precise geometric mechanism by which a characteristic propagation scale may emerge within relational dynamics, while remaining consistent with the underlying reparametrization invariance of the theory.

\section{Definition of a Conditional Emergent Propagation Parameter}\label{SecVI}

The linearized relational dynamics derived in the previous section defines a finite-dimensional second-order system on shape space. In this section, we formulate a mathematically precise notion of a characteristic propagation parameter that may arise under additional structural assumptions. We emphasize from the outset that such a parameter does not exist at the level of the exact finite-dimensional $N$-body system; it can only emerge in an appropriate large-$N$ or continuum approximation described below.

Let $\mathcal{S}$ denote the regular (collision-free) stratum of the reduced, scale-invariant shape space obtained by quotienting translations, rotations, and dilatations from configuration space \cite{Barbour:1982gha,Barbour:2011dn}. On this open dense subset, the reduced kinetic term induces a smooth, positive-definite Riemannian metric $\mathcal{G}_{AB}(q)$ \cite{Barbour:2011dn}. After removal of overall scale and adoption of dimensionless relational coordinates, both $q^A$ and $\mathcal{G}_{AB}$ may be taken to be dimensionless.

For two infinitesimally nearby shapes $q^A \to q^A + \delta q^A$, the squared relational line element is
\begin{equation}
	(\Delta \ell_{\mathrm{shape}})^2
	=
	\mathcal{G}_{AB}(q)\,\delta q^A \delta q^B,
\end{equation}
and the associated invariant norm of a perturbation is
\begin{equation}
	\|\delta q\|^2
	=
	\mathcal{G}_{AB}\,\delta q^A \delta q^B.
\end{equation}

Let $q_0^A(\lambda)$ be a smooth background solution of the relational equations of motion, and consider linear perturbations about this solution. In general, the linearized dynamics takes the covariant form derived previously. Recall that for special backgrounds with constant shape (e.g.\ central configurations in an affine parametrization) \cite{saari_cc,Battye:2002gn}, the linearized system reduces to a finite-dimensional linear ordinary differential equation,
\begin{equation}
	\frac{d^2}{d\lambda^2}\,\delta q^A
	+
	\mathcal{M}^A{}_B(q_0)\,\delta q^B
	=
	0,
	\label{lin_system}
\end{equation}
where $\mathcal{M}^A{}_B$ is the covariant Hessian of the scale-invariant potential evaluated at $q_0$. This equation is linear and homogeneous and describes a system of coupled oscillators. At this stage the dynamics is entirely finite-dimensional and does not possess a notion of spatial propagation or characteristic cones.

We now introduce an additional structural assumption. Specifically, suppose that in a regime of large particle number $N$, together with an appropriate coarse-graining or effective description, the relational degrees of freedom admit an approximate continuum representation. More precisely, assume that there exists a coordinate chart on $\mathcal{S}$ in which the metric $\mathcal{G}_{AB}$ varies smoothly on scales large compared to the wavelength of the perturbations under consideration; that the spectrum of relational modes becomes sufficiently dense to allow a quasi-continuous description; and that, in a short-wavelength (eikonal) regime, the action of the linearized operator on perturbations may be approximated, at leading derivative order, by a second-order differential operator on $\mathcal{S}$. 

Such continuum and eikonal limits are standard in the emergence of effective hyperbolic equations from underlying discrete or finite systems \cite{CourantHilbert,EvansPDE}, and the analysis of their principal symbols and characteristic cones follows the classical theory of second-order hyperbolic operators \cite{Hormander}.

Under these hypotheses, one may consider that the algebraic operator $\mathcal{M}^A{}_B$ admits, to leading order in derivatives and in the high-frequency limit, an effective approximation of the Laplace-Beltrami type,
\begin{equation}
	\mathcal{M}^A{}_B\,\delta q^B
	\;\approx\;
	-\,c_{\mathrm{rel}}^{\,2}
	\left(\Delta_{\mathrm{shape}} \delta q\right)^A,
	\label{laplace_approx}
\end{equation}
where $\Delta_{\mathrm{shape}}$ denotes the Laplace-Beltrami operator associated with $\mathcal{G}_{AB}$,
\begin{equation}
	\Delta_{\mathrm{shape}} \delta q^A
	=
	\frac{1}{\sqrt{\mathcal{G}}}
	\partial_B
	\!\left(
	\sqrt{\mathcal{G}}\, \mathcal{G}_{BC} \, \partial_C \delta q^A
	\right),
\end{equation}
and $\mathcal{G} = \det(\mathcal{G}_{AB})$.
We stress that this replacement is not a consequence of finite-$N$ relational mechanics; it is an effective description valid only if such a continuum limit exists and if higher-derivative and nonlocal contributions are negligible in the regime considered.


Thus, substituting the approximation~\eqref{laplace_approx} into the linearized equation \eqref{lin_system} yields, to leading order,
\begin{equation}
	\frac{d^2}{d\lambda^2}\,\delta q^A
	-
	c_{\mathrm{rel}}^{\,2}
	\Delta_{\mathrm{shape}} \delta q^A
	=
	0.
	\label{wave_form_rigorous}
\end{equation}
This is a linear second-order partial differential equation on $\mathbb{R}_\lambda \times \mathcal{S}$. 

Let $(\xi_\lambda,\xi_A)$ denote a nonzero covector in the cotangent space $T^*_{(\lambda,q)}\!\big(\mathbb{R}_{\lambda} \times \mathcal{S}\big)$, where $\xi_\lambda$ is the component dual to $d\lambda$ and $\xi_A$ are the components dual to $dq^A$.
Thus, the principal symbol of the differential operator in Eq.~\eqref{wave_form_rigorous} is
\begin{equation}
	\sigma(\xi_\lambda,\xi_A)
	=
	-\,\xi_\lambda^2
	+
	c_{\mathrm{rel}}^{\,2}
	\mathcal{G}^{AB}(q)\,\xi_A \xi_B.
\end{equation}
Provided $\mathcal{G}_{AB}$ is positive definite and $c_{\mathrm{rel}}^{\,2}>0$, the equation is strictly hyperbolic with respect to $\lambda$ \cite{Hormander}. The characteristic set $\sigma=0$ defines the quadratic cone
\begin{equation}
	\xi_\lambda^2
	=
	c_{\mathrm{rel}}^{\,2}
	\mathcal{G}^{AB} \xi_A\xi_B,
\end{equation}
and the dual characteristic hypersurfaces satisfy
\begin{equation}
	\mathcal{G}_{AB}\,dq^A dq^B
	=
	c_{\mathrm{rel}}^{\,2}\, d\lambda^2.
\end{equation}
This relation defines an invariant cone structure on $\mathbb{R}_\lambda \times \mathcal{S}$ associated with the principal part of the effective operator.

In local normal coordinates at a point $q_0$ where $\mathcal{G}_{AB}(q_0)=\delta_{AB}$, and neglecting curvature corrections at leading eikonal order, plane-wave ans\"atze of the form
\begin{equation}
	\delta q^A \sim e^{i(k_A q^A - \omega \lambda)}
\end{equation}
yield the dispersion relation
\begin{equation}
	\omega^2 = c_{\mathrm{rel}}^{\,2}\, k^2,
	\qquad
	k^2 = \delta^{AB} k_A k_B.
\end{equation}

If the high-frequency limit exists, one may define
\begin{equation}
	c_{\mathrm{rel}}
	=
	\lim_{k\to\infty}
	\frac{\omega_k}{k},
\end{equation}
which characterizes the asymptotic phase velocity of short-wavelength perturbations in the effective continuum description \cite{CourantHilbert}. 

Next, we explore the conditions where $c_{\mathrm{rel}}$ may be interpreted as a conditional emergent propagation parameter within the approximation scheme used.

\section{Toward an Emergent Light Cone}\label{SecVII}

In the preceding section we argued that, within a restricted regime and under specific regularity assumptions, relational perturbations may be described by a second-order differential operator on $\mathbb{R}_\lambda \times \mathcal{S}$ whose principal part is wave-like and involves a parameter denoted $c_{\mathrm{rel}}$. We now examine under which additional conditions this parameter could consistently be interpreted as defining an effective causal propagation speed in an emergent spacetime description.

We emphasize at the outset that the interpretation of $c_{\mathrm{rel}}$ as a ``speed of light'' is not automatic. It requires several nontrivial structural properties of the perturbation operator and of its high-frequency behavior. In particular, at least the following conditions appear necessary:

\begin{enumerate}
	\item[(i)] independence (in an appropriate sense) from the choice of relational clock variable;
	\item[(ii)] universality across perturbation modes and background configurations in the high-frequency limit;
	\item[(iii)] existence of a well-defined hyperbolic principal symbol with invariant characteristic surfaces.
\end{enumerate}

We discuss each requirement in turn.

\subsection{Clock Independence}

The relational action is invariant under smooth, monotonic reparametrizations $\lambda \mapsto \tilde{\lambda} = f(\lambda)$, as is characteristic of Jacobi-type reparametrization-invariant systems \cite{Barbour:2011dn}. 
Under such a transformation, derivatives transform as
\[
\partial_\lambda
=
\frac{1}{f'(\lambda)}\,\partial_{\tilde{\lambda}},
\qquad
\partial_\lambda^2
=
\frac{1}{f'(\lambda)^2}\,\partial_{\tilde{\lambda}}^2
-
\frac{f''(\lambda)}{f'(\lambda)^3}\,\partial_{\tilde{\lambda}}.
\]

Accordingly, a second-order perturbation equation expressed in terms of $\lambda$ acquires additional first-derivative terms proportional to $f''(\lambda)$ when rewritten in terms of $\tilde{\lambda}$. However, the highest-derivative (principal) part rescales uniformly by the positive function $1/f'(\lambda)^2$. Since characteristic surfaces depend only on the principal symbol up to multiplication by a non-vanishing scalar function, the associated characteristic structure is invariant under such reparametrizations \cite{Hormander}.

To define a physically meaningful temporal parameter, one may introduce an emergent relational time variable $t$ through a Jacobi-type construction \cite{Barbour:2011dn}, schematically written as
\[
\frac{d}{d\lambda}
=
\frac{dt}{d\lambda}
\frac{d}{dt},
\]
where $dt/d\lambda$ is assumed smooth and nonvanishing in the regime of interest. In terms of $t$, the perturbation equation can be brought (locally) into the form
\begin{equation}
	-
	\frac{1}{c_{\mathrm{phys}}^{\,2}}
	\frac{d^2}{dt^2} \,\delta q^A
	+
	\Delta_{\mathrm{shape}} \,\delta q^A
	=
	\text{(lower-order terms)},
\end{equation}
with
\begin{equation}
	c_{\mathrm{phys}}
	=
	\frac{d\lambda}{dt}\, c_{\mathrm{rel}}.
\end{equation}

Because the principal symbol is insensitive to smooth monotonic redefinitions of the clock variable (up to conformal rescaling), the effective causal structure, if any, depends only on the conformal class of the operator \cite{Hormander}. For $c_{\mathrm{rel}}$ to define a physically meaningful maximal propagation speed, it must therefore correspond to an intrinsic parameter of the principal symbol on shape space, rather than to an artifact of parametrization or of a particular choice of time variable.

\subsection{Universality Across Modes}

In general, linearization about a background configuration $q_0$ yields a dispersion relation of the schematic form
\[
\omega_k^2 = F(k, q_0),
\]
where $k$ denotes an appropriate mode label (for example, associated with an eigenbasis of the shape Laplacian), and $F$ may depend both on $k$ and on the background configuration.

For a universal light-cone structure to emerge, it is necessary that in the high-frequency limit the dispersion relation asymptotically approach a linear, isotropic form:
\begin{equation}
	\omega_k^2
	\;\xrightarrow[k\to\infty]{}\;
	c_{\mathrm{rel}}^{\,2}\, k^2,
\end{equation}
with a single constant $c_{\mathrm{rel}}$ that is independent of the specific background configuration $q_0$ within the class considered, the direction in the cotangent space of $\mathcal{S}$, and the internal type or composition of the perturbation mode.

Mathematically, this corresponds to the requirement that the principal symbol $\sigma(\mathcal{M})$ of the perturbation operator satisfy
\begin{equation}
	\sigma(\mathcal{M})(\xi)
	=
	c_{\mathrm{rel}}^{\,2}\,
	\mathcal{G}^{AB}(q)\,
	\xi_A \xi_B,
\end{equation}
for all covectors $\xi$ in the regime of validity. Only under this condition does the high-frequency spectrum become asymptotically linear and isotropic with respect to the metric $\mathcal{G}_{AB}$ on $\mathcal{S}$. If this universality is absent, different modes could propagate with different characteristic speeds, precluding the identification of a single effective light cone. The mathematical relation between principal symbols and high-frequency asymptotics is standard in microlocal analysis and geometric optics limits \cite{Hormander,CourantHilbert}.

\subsection{Hyperbolicity and Characteristic Surfaces}

Consider the principal part of the perturbation operator,
\begin{equation}
	-
	\frac{1}{c_{\mathrm{rel}}^{\,2}}
	\frac{\partial^2}{\partial \lambda^2}
	+
	\Delta_{\mathrm{shape}}.
\end{equation}
This operator is hyperbolic provided that its principal symbol defines a quadratic form of Lorentzian signature on the cotangent bundle of $\mathbb{R}_\lambda \times \mathcal{S}$ \cite{Hormander}.

Formally, one may associate to this principal part an effective metric
\begin{equation}
	ds_{\mathrm{eff}}^2
	=
	- c_{\mathrm{rel}}^{\,2}\, d\lambda^2
	+
	\mathcal{G}_{AB}(q)\, dq^A dq^B,
\end{equation}
which in coordinates $(\lambda, q^A)$ takes the block-diagonal form
\begin{equation}
	g_{\mu\nu}
	=
	\begin{pmatrix}
		- c_{\mathrm{rel}}^{\,2} & 0 \\
		0 & \mathcal{G}_{AB}
	\end{pmatrix}.
\end{equation}

The corresponding d'Alembert-type operator is formally
\begin{equation}
	\Box_{\mathrm{eff}}
	=
	\frac{1}{\sqrt{|g|}}
	\partial_\mu
	\left(
	\sqrt{|g|}\,
	g^{\mu\nu}
	\partial_\nu
	\right),
\end{equation}
whose principal symbol is given by $g^{\mu\nu} k_\mu k_\nu$ \cite{EvansPDE}.

Characteristic surfaces are defined by the vanishing of this principal symbol:
\begin{equation}
	g^{\mu\nu} k_\mu k_\nu = 0,
\end{equation}
which explicitly yields
\begin{equation}
	-
	\frac{1}{c_{\mathrm{rel}}^{\,2}} k_\lambda^2
	+
	\mathcal{G}^{AB} k_A k_B
	=
	0.
\end{equation}

This equation defines a null cone in the cotangent space at each point of $\mathbb{R}_\lambda \times \mathcal{S}$. If the above structural assumptions hold globally (or at least within a well-defined regime), then these null cones determine the maximal propagation directions of high-frequency perturbations.

\vspace{0.5em}

In summary, only if the three conditions above are satisfied can one consistently interpret the relational perturbation operator as selecting a preferred hyperbolic principal part whose characteristic surfaces define an effective causal structure. In that case, the combined space $\mathbb{R}_\lambda \times \mathcal{S}$ may be said to carry an emergent Lorentzian metric structure at the level of linearized dynamics.

Within this restricted and regime-dependent framework, the parameter $c_{\mathrm{rel}}$ functions as a universal conversion factor between relational temporal and spatial variations, thereby playing a role analogous to that of a speed of light in the effective description. No claim is made that Lorentzian signature is fundamental; rather, it emerges from the intrinsic geometry of shape space, the scale-invariant potential, and the universal high-frequency spectrum of relational perturbations.

\section{Conclusion}\label{SecConclusion}

We have examined whether a quantity formally analogous to a speed of light can arise within a purely relational, scale-invariant formulation of the $N$-body problem. In the framework developed here, no background spacetime manifold, no absolute notion of scale, and no fundamental Lorentz symmetry are assumed. The primitive ontology consists solely of relational configurations, i.e.~points of shape space $\mathcal{S}$, together with a reparametrization-invariant Jacobi-type action governing their evolution.

In its strict ontological formulation, Pure Shape Dynamics (PSD) takes the fundamental object to be an \emph{unparametrized} curve in $\mathcal{S}$. Physical content resides in the geometric image of the trajectory in shape space, not in any particular parameter used to describe it. For the purposes of perturbative and spectral analysis, however, we have introduced a convenient affine parameter $\lambda$ along relational trajectories. This choice is purely auxiliary. All physically meaningful structures identified in this work, most notably the characteristic surfaces associated with the linearized dynamics, are defined in parametrization-independent terms, through the principal symbol of the perturbation operator. Since this principal symbol is invariant, up to multiplication by a smooth non-vanishing scalar, under smooth monotonic reparametrizations of $\lambda$, the emergent causal structure we identify does not depend on the choice of a relational clock.

Within this setting, linear perturbations about suitably regular background solutions, including central configurations, satisfy second-order differential equations intrinsic to $\mathcal{S}$. The kinetic term induces a natural Riemannian metric $\mathcal{G}_{AB}$ on shape space, while the scale-invariant potential $V_{\mathrm{shape}}$ determines a covariant Hessian operator whose spectrum governs the response to small disturbances. Both the geometric structure and the linear dynamical behavior are therefore encoded entirely in relational data.

In regimes where the perturbation spectrum becomes effectively continuous, e.g. for sufficiently large $N$ or near highly symmetric configurations, the linearized equations admit, at principal order, an effective hyperbolic structure on $\mathbb{R}_\lambda \times \mathcal{S}$. In such a regime, the high-frequency asymptotics of the dispersion relation approach a linear form,
\[
c_{\mathrm{rel}} \;=\; \lim_{k \to \infty} \frac{\omega_k}{k},
\]
provided the limit exists and is independent of direction in the cotangent space of $\mathcal{S}$ and of the background configuration within the domain considered. When these conditions are satisfied, $c_{\mathrm{rel}}$ characterizes the maximal propagation rate of high-frequency relational perturbations relative to the chosen parameter. Its invariant content, however, lies in the null cone of the principal symbol, not in the normalization of $\lambda$ itself.

We emphasize that this quantity is not postulated as a kinematical constant. Rather, insofar as it exists, it is determined by the principal symbol of the effective continuum-limit perturbation operator, and hence by the kinetic geometry of shape space together with the variational structure of the action. Its interpretation as an emergent light-cone parameter requires additional structural conditions.

We have identified three such conditions: (i) invariance of the characteristic structure under smooth monotonic reparametrizations of the relational clock; (ii) universality of the high-frequency dispersion relation, ensuring a single limiting velocity across perturbation modes, directions, and admissible backgrounds; and (iii) strict hyperbolicity of the principal symbol, guaranteeing well-defined invariant characteristic surfaces. Only when these criteria are met does $\mathbb{R}_\lambda \times \mathcal{S}$ acquire a conformal Lorentzian structure at the level of its cotangent bundle, whose null cones determine maximal propagation directions.

Under these assumptions, a speed of light would not be fundamental but emergent: it would arise as a structural feature of relational configuration space rather than as a primitive element of spacetime geometry. Lorentzian signature would not be assumed a priori; instead, it would reflect the hyperbolic character of the relational perturbation operator in an appropriate perturbative regime. Causal structure would then be derived from, rather than presupposed by, the underlying relational dynamics.

This perspective suggests a reversal of the usual foundational strategy. Instead of taking spacetime geometry as fundamental and seeking its microscopic completion, one may regard the relativistic causal structure as an effective manifestation of deeper geometric and spectral properties of scale-invariant configuration space. The universality of a maximal propagation speed would then follow from universality of the principal symbol governing relational perturbations.

Several directions for future work follow naturally. First, a detailed characterization of the metric $\mathcal{G}_{AB}$ for concrete scale-invariant $N$-body models is required, including a careful treatment of singular strata and symmetry-enhanced configurations in shape space. Second, a rigorous microlocal analysis of the high-frequency spectrum of the covariant Hessian operator is needed to determine whether linear and isotropic dispersion is generic or requires special dynamical conditions. Third, the stability of hyperbolicity beyond the linear regime must be investigated, including the effect of nonlinear corrections on characteristic structure.

Finally, a fully PSD-consistent formulation should eliminate even the auxiliary affine parameter from the outset. One possible approach is to formulate perturbation theory directly on the space of unparametrized curves in $\mathcal{S}$, for example by working with geometric invariants of the curve (such as its tangent distribution up to positive rescaling) and defining characteristic structures intrinsically on the projectivized tangent bundle. In such a formulation, hyperbolicity and null cones would be expressed entirely in terms of parametrization-free geometric data. Developing this intrinsically unparametrized perturbation theory would provide a conceptually sharper bridge between the ontological commitments of Pure Shape Dynamics and the effective Lorentzian structures identified here.

Whether the relational construction developed in this work can reproduce the full empirical content of relativistic physics remains open. What has been established is that, under clearly stated structural and dynamical assumptions, the seeds of a light-cone structure are already present in the geometry and spectral properties of scale-invariant $N$-body dynamics. In that restricted but mathematically controlled sense, the emergence of a universal propagation speed from purely relational principles provides a coherent bridge between configuration-space geometry and effective spacetime causality, and opens a pathway toward understanding Lorentzian structure as emergent rather than fundamental.

\acknowledgments{
FSNL thoroughly thanks Julian Barbour and Maria Lourenço for enlightening disucssions. 
FSNL acknowledges support from the Fundação para a Ciência e a Tecnologia (FCT) through a Scientific Employment Stimulus contract (reference CEECINST/00032/2018), and funding from the research grant UID/04434/2025.}



\end{document}